\documentclass[aps, twocolumn]{revtex4}
\usepackage{amsmath, amsthm, amssymb}
\usepackage{epsfig}
\usepackage{graphicx}

\begin{document}

\title
  {Thermal decomposition of a honeycomb-network sheet -- A
Molecular Dynamics simulation study}

 \author{J. Paturej$^{1,2}$\footnote{Corresponding author.
Email:~jpaturej@univ.szczecin.pl}, H. Popova$^{3}$, A. Milchev$^{1,3}$ and T.A.
Vilgis$^1$ }
 \affiliation{$^1$ Max Planck Institute for Polymer Research, 10 Ackermannweg,
 55128 Mainz, Germany\\
 $^2$ Institute of Physics, University of Szczecin, Wielkopolska 15,
 70451 Szczecin, Poland\\
 $^3$ Institute for Physical Chemistry, Bulgarian Academy of Sciences, 1113
 Sofia, Bulgaria}

\begin{abstract}
The thermal degradation of a graphene-like two-dimensional
honeycomb membrane
with bonds undergoing temperature-induced scission is studied by means of
Molecular Dynamics simulation using Langevin thermostat. We demonstrate that at
lower temperature the
probability distribution of breaking bonds is highly peaked at the rim of the
membrane sheet  whereas at higher temperature bonds break at
random everywhere in the hexagonal flake. The mean breakage time $\tau$ is found
to decrease with the total number of network nodes $N$ by a power law $\tau
\propto N^{-0.5}$ and reveals an Arrhenian dependence on temperature $T$.
Scission times are themselves exponentially distributed. The fragmentation
kinetics of the average number of clusters can be described by first-order
chemical reactions between network nodes $n_i$ of different coordination. The
distribution of fragments sizes evolves with time elapsed from initially a
$\delta$-function through a bimodal one into a single-peaked again at late
times. Our simulation results are complemented by a set of $1^{st}$-order
kinetic differential equations for $n_i$ which can be solved exactly and
compared to data derived from the computer experiment, providing deeper insight
into the thermolysis mechanism.
\end{abstract}

\maketitle

\section{Introduction} \label{sec:intro}

Thermal degradation and stabilization of polymer systems has been a
long-standing focus of research from both practical and fundamental viewpoints
\cite{Allen}. Plastic waste disposal has grown rapidly to ecological menace
prompting researchers to investigate plastic recycling by degradation as an
alternative \cite{Madras}. On the other hand, degradation of polymers and other
high molecular weight materials in different environments is usually a major
limiting factor in their application. Thermal degradation (or, {\em
thermolysis}) plays a decisive role in the design of flame-resistant
polyethylene and other plastic materials \cite{Nyden}. Another interesting
aspects for applications include reversible polymer networks
\cite{Sijbesma1,Sijbesma2}, and most notably, graphene, as a ''material of the
future'' that shows unusual thermomechanical properties \cite{Peeters,Aluru}.
Recently, with the rapidly growing perspective of exploiting bio-polymers as
functional materials \cite{Schulten,Han}, the stability of  such materials has
become an issue of primary concern \cite{Lindahl,Sinha} as, e.g., that of
double-stranded polymer decomposition \cite {Metanomski}.

Most theoretical investigations of polymer degradation have focused on
determining the rate of change of average molecular weight
\cite{Jellinek,Ballauff,Ziff,Cheng,Nyden2,Doerr,Wang,Hathorn,Doruker,Flegg}.
The main assumptions of the theory are that each link in a long chain molecule
has equal strength and equal accessibility, that they are broken at random, and
that the probability  of rupture is proportional to the number of links present.
Therefore, all of the afore-mentioned studies investigate exclusively the
way in which the distribution of bond rupture probability along the polymer
backbone affects the fragmentation kinetics and the distribution of fragment
sizes as time elapses. In a recent study \cite{Paturej,PaturejEPL}, using
Molecular
Dynamics (MD) simulation with a Langevin thermostat, we observed a rather
complex interplay between the polymer chain dynamics and the resulting bond
rupture probability. A major role in this was attributed to the one-dimensional
(1D) topological connectivity of the linear polymer.
Significant change in rupture kinetics is observed while polymer architecture
is tuned as in the case of thermolysis of adsorbed bottle-brushes
\cite{bb1,bb2}.

Understanding the interplay between elastic and fracture properties  is even more
challenging and
important
in the
case  of 2D polymerized networks (elastic-brittle sheets).
A prominent example of biological microstructure is {\em spectrin},
the  red blood cell membrane skeleton, which reinforces the cytoplasmic face of
the membrane. In erythrocytes, the membrane skeleton enables it to undergo
large extensional deformations while maintaining the structural integrity of the
membrane. A number of studies, based on continuum- \cite{Skalak}, percolation-
\cite{Srolovitz,Saxton,Seifert}, or molecular level \cite{Monette,Dao}
considerations of the mechanical breakdown of this network, modeled as a
triangular lattice of spectrin tetramers, have been reported so far. Another
example concerns the thermal stability of isolated graphene nanoflakes.
It has been
investigated recently by Barnard and Snook \cite{Snook} using {\em ab initio}
quantum mechanical techniques whereby it was noted that the problems ``has
been overlooked by most computational and theoretical studies''. Many of these
studies can be viewed in a broader context as part of the problem of thermal
decomposition of gels \cite{Argyropoulos}, epoxy resins \cite{Rico,Zhang} and
other 3D networks, studied both experimentally \cite{Argyropoulos,Rico,Zhang}, and by
means of simulations \cite{Kober} as in the case of Poly-dimethylsiloxane (PDMS).
In most of these cases, however, mainly a stability analysis is carried out
whereas still little is known regarding the collective mechanism of
degradation, the dependence of rupture time on system size,
as well as the decomposition  kinetics, especially as far as (2D) polymer
network sheets are concerned. It is also interesting from the standpoint of
basic physics to compare the degradation process to the one taking place in
linear polymers where considerable progress has been achieved recently
\cite{Paturej}. Therefore, in the present work we extend our investigations to
the case of (2D) polymer network sheets, embedded in 3D-space, and study as a
generic example the thermal degradation of a suspended membrane with honeycomb
orientation. 

The paper is organized as follows: after a brief introduction, we sketch our
model in Sec.~\ref{sec:model}.
In Sec.~\ref{sec_MD_results} we present our simulation results, that is, the
distribution of bond scission rates over the membrane surface, the Mean First
Breakage Time (MFBT) of a bond depending on membrane size and temperature,
the distribution of recombination events -  Sec.~\ref{subsec_tau}, and the
temporal evolution of the fragmentation process - Sec.~\ref{subsec_fragment}. We
also develop a theoretical scheme based on a set of $1^{st}-$order kinetic
differential equations, describing the time variation of the number of network
nodes, connected by a particular number of bonds to neighboring nodes -
 Sec.~\ref{sec_diff_eq}. We demonstrate that the analytical solution of such
system provides a faithful description of the fragmentation kinetics. In
Sec.~\ref{sec_summary} we conclude with a brief discussion of our main findings.

\section{METHODS} \label{sec:model}

\subsection{Model}

\begin{figure}[ht]
\begin{center}
\includegraphics[scale=0.31]{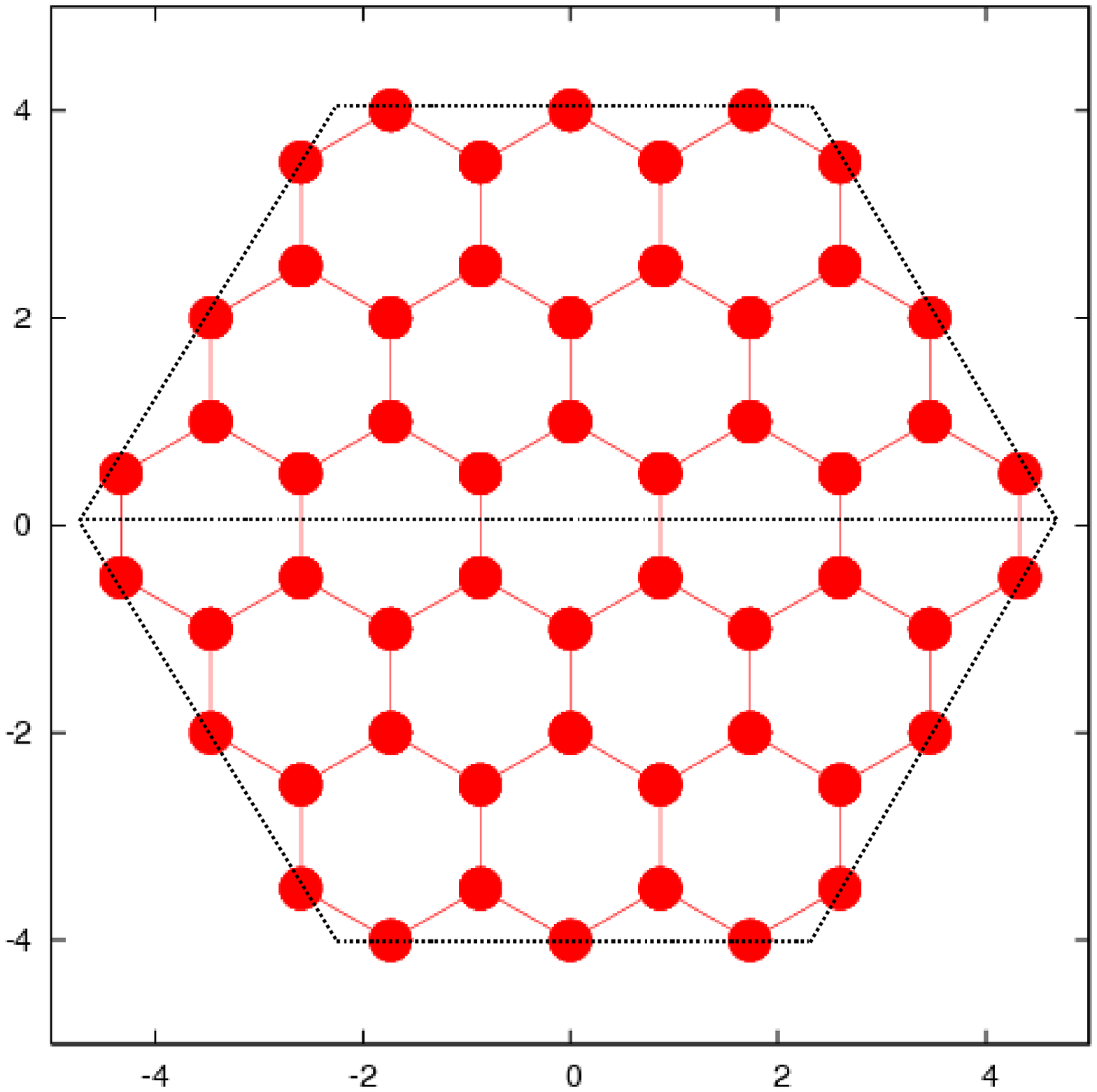}
\hspace*{0.8cm}
\includegraphics[scale=0.35]{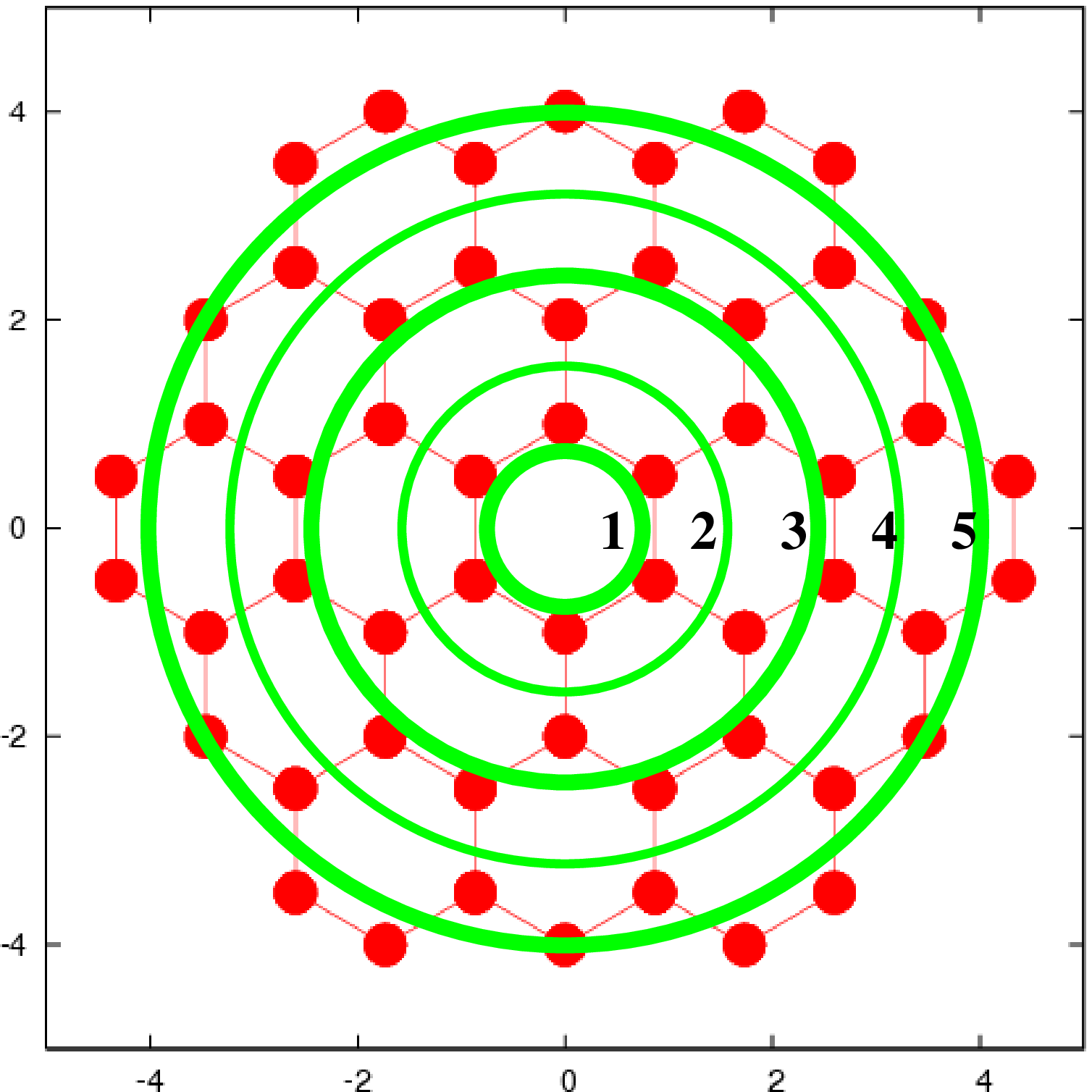}
\caption{[upper panel] A model of a membrane with honeycomb structure that
contains a total of $N=54$ beads and has linear size $L=3$ ($L$ is the number of
beads or hexagonal cells on the edge of the membrane). [lower panel] An example
of subdivision of beads and bonds, composing a membrane with $L=3$, into
subgroups (``circles''). The total number of circles $C$ in the membrane of
linear size $L$ is $C=2L-1$. \label{fig_ModelMembrane}}
\end{center}
\end{figure}

We study a coarse-grained model of honeycomb membrane embedded in
three-dimensional (3D) space. 
In this investigation  we consider generally {\em symmetric} hexagonal
membranes ({\em flakes}) (Fig.~\ref{fig_ModelMembrane}). In a very few cases we
also discuss fracture of a {\it ribbon} shape membranes.
The membrane flake consists of $N$ spherical particles
(beads, monomers) of diameter $\sigma$ connected in a honeycomb lattice
structure whereby each monomer is bonded with three nearest-neighbors
except of the monomers on the membrane edges which have only two bonds (see
Fig.~\ref{fig_ModelMembrane} [upper panel]). The total number of monomers $N$ in
such a membrane is $N=6L^2$ where by $L$ we denote the number of monomers (or
hexagon cells) on the edge of the membrane (i.e., $L$ characterizes the linear
size of the membrane). There are altogether $N_{bonds}=(3N-6L)/2$ bonds in the
membrane. 

We find it appropriate to divide the two-dimensional membrane network so that
all the beads and bonds are distributed into different subgroups presented by
concentric ``circles'' with consecutive numbers (see
Fig.~\ref{fig_ModelMembrane} [upper panel]) proportional to the radial distance
from the membrane center. To {\em odd} circle numbers thus belong beads and
bonds that are nearly tangential to the circle. {\em Even} circles contain no
beads and only radially oriented bonds (shown to cross the circle in
Fig.~\ref{fig_ModelMembrane}. The total number of circles $C$ in the membrane
of linear size $L$ is found to be $C=(2L-1)$. We use this example of dividing
the beads and the bonds composing the membrane in order to represent our
simulation
results in appropriate way which relates them to their relative proximity to
membrane's periphery.

\subsection{Potentials}

The nearest-neighbors in the membrane are connected to each other by "breakable
bonds" described by a Morse potential, where $r$ is a distance between the
monomers,
\begin{equation}
U_{\text{M}}(r) = \epsilon_M \lbrace 1 - \exp[-\alpha(r-r_{\text{min}})] \rbrace
^{2}
\label{eq_U_MORSE}
\end{equation}
$\alpha=1$ is a constant that determines bond elasticity and $r_{\text{min}}
= 1$ is the equilibrium bond length. The dissociation energy of a given bond is
$\epsilon_M=1$, measured in units of $k_BT$, where $k_B$ denotes the Boltzmann
constant and $T$ is the temperature. The minimum of this potential occurs at
$r=r_{\text{min}},\;~ U_{\text{Morse}}(r_{\text{min}}) = 0$. The maximal
restoring force of the Morse potential, $f_{\text{max}} = -dU_{\text{M}}/dr =
\alpha \epsilon_M /2$, is reached at the inflection point,
$r_{\mbox{\tiny inflex}} =
r_{\text{min}} +
\alpha^{-1} \ln(2)\approx 2.69$.
This force $f_{\text{max}}$ determines the maximal tensile
strength of the membrane.
Stretching of the bond beyond $r_{\mbox{\tiny inflex}}$ means potentially a
scission of that
bond as far as the restoring force declines rapidly with bond length even
though there is a chance for a recombination.
Therefore we take as a criterion for  breaking a
bond its expansion to $r_h=5$ beyond which practically a bond
recombinantion is ruled out.
Since $U_{\text{M}}(0) \approx 2.95$, the Morse
potential, Eq.~(\ref{eq_U_MORSE}), is only weakly repulsive and beads could
partially penetrate one another at $r < r_{\text{min}}$. Therefore, in order to
allow properly for the excluded volume interactions between bonded monomers, we
take the bond potential as a sum of $U_{\text{M}}(r)$ and the so called
Weeks-Chandler-Anderson (WCA) potential, $U_{\text{WCA}}(r)$, (i.e., the shifted
and truncated repulsive branch of the Lennard-Jones potential),
\begin{eqnarray}
U_{\text{WCA}}(r) = \begin{cases}
4\epsilon \left[ \left( \frac{\sigma}{r} \right) ^{12} - \left( \frac{\sigma}{r}
\right) ^{6} \right] + \epsilon,  & \text{for}~~ r \leq 2^{1/6}\sigma \\
0,  & \text{for}~~ r > 2^{1/6}\sigma
\end{cases}
\label{aaa}
\end{eqnarray}
with parameter $\epsilon=1$ and monomer diameter $\sigma=2^{-1/6} \approx 0.89$
so that the minimum of the WCA potential coincides with the minimum of the
Morse potential. Thus, the length scale is set by the parameter $r_{\text{min}}
= 2^{1/6}\sigma = 1$. The nonbonded interactions between monomers are taken
into account by means of the WCA potential, Eq.~(\ref{aaa}). The nonbonded
interactions in our model correspond to good solvent conditions whereas the
bonded interactions make the bonds in our model breakable so they undergo
scission at sufficiently high $T$.

We have been anxious to
emphasize the common features of failure in materials with similar architecture
but largely varying elasticity properties, e.g., from $1000$~GPa graphene's
Young modulus \cite{Aluru} compared to $4\times 10^{-3}$~GPa for
spectrin \cite{Dao}. Putting the value of a Kuhn segment to $\sigma=1.44$~\AA~and
taking
the thermal energy $k_BT=4\times 10^{-21}$~J at $T=300$~K, we get from our
simulation 
\cite{graphene_force} 
a Young modulus $0.03$~GPa
which is ranged between typical values for rubber-like materials
$0.01$--$0.1$~GPa.

\subsection{MD algorithm}

In our MD simulation we use Langevin dynamics, which describes the Brownian
motion of a set of interacting particles whereby the action of the solvent is
split into slowly evolving viscous (frictional) force and a rapidly fluctuating
stochastic (random) force. The Langevin equation of motion is the following:
\begin{equation}
m\overrightarrow{\dot{v}_i}(t) = \overrightarrow{F}_i(t) - m \gamma
\overrightarrow{v_i}(t) + \overrightarrow{R}_i(t)
\label{eq_Langevin}
\end{equation}
where $m$ denotes the mass of the particles which is set to $m=1$,
$\overrightarrow{v}_i$ is the velocity of particle $i$, $\overrightarrow{F}_i =
(\overrightarrow{F}_\textmd{M} + \overrightarrow{F}_\textmd{WCA})_i$ is the
conservative force which is a sum of all forces exerted on particle $i$ by other
particles in the system, $\gamma$ is the friction coefficient and
$\overrightarrow{R}_i$ is the three dimensional vector of random force acting on
particle $i$. The random force $\overrightarrow{R}_i$, which represents the
incessant collision of the monomers with the solvent molecules, satisfies the
fluctuation-dissipation theorem $\langle R_{i\alpha}(t) R_{j\beta}(t') \rangle =
2 \gamma k_B T \delta_{ij} \delta_{\alpha\beta} \delta(t-t')$ where the
symbol $\langle \ldots \rangle$ denotes an equilibrium average and the greek-letter
subscripts refer to the $x$, $y$ or $z$ components. The friction coefficient
$\gamma$ of the Langevin thermostat is generally set to $\gamma=0.25$ and only
in very few cases to $10$.  The integration
step is $0.002$ time units (t.u.) and the time is measured in units of
$r_{\text{min}} \sqrt{m/\epsilon_M}$. We emphasize at this point that in our
coarse-grained modeling no explicit solvent particles are included. In this work
the velocity-Verlet algorithm is used to integrate the equations of motion.

\begin{figure}[!h]
\begin{center}
\includegraphics[scale=0.41]{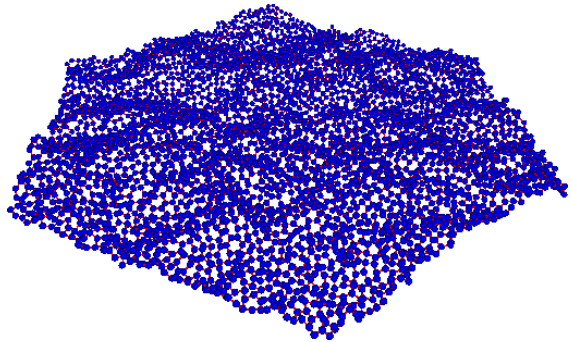}
\caption{A snapshot of a typical conformation of an intact membrane with $L =
30$ containing $5400$ monomers after equilibration. Characteristic ripples are
seen to
cross the surface. \label{fig_EquilMembrane}}
\end{center}
\end{figure}

Our MD simulations are carried out in the following order. First, we prepare an
equilibrated membrane conformation, starting with a fully flat configuration
shown schematically in Fig.~\ref{fig_ModelMembrane}, where each bead in the
network is separated by a
distance $r_{\text{min}}=1$ equal to the equilibrium separation of the bond
potential $(U_{\text{M}}+U_{\text{WCA}})$ [see Eq.~(\ref{eq_U_MORSE}), and
Eq.~(\ref{aaa})]. 
Then we start the simulation with this prepared conformation
and let the membrane equilibrate in the heat bath  for a very long period of
time ($\approx 10^7$ integration steps) at a temperature low enough
that the membrane stays intact, Fig.~\ref{fig_EquilMembrane}, (this
equilibration is done in order to prepare different starting conformations for
each simulation).
Then the temperature is raised to the working temperature and the membrane is
equilibrated for 20 MD t.u.~($10^4$ integration steps) which interval was
found as sufficient to establish equipartition (uniform distribution of the
temperature throughout the membrane).
The time then is set to zero and we continue the simulation of this
 membrane conformation to examine the thermal scission of
the
bonds. We measure the elapsed time $\tau$ until the first bond rupture occurs
and repeat the above procedure for a large number of events ($10^3$--$10^4$) so
as to sample the stochastic nature of rupture and to determine the mean $\langle
\tau \rangle$ which we refer to as the Mean First Breakage Time. In the
course of simulation we also calculate properties such as the probability
distribution of breaking bonds regarding their position in the membrane (a
rupture probability histogram), the probability distribution of the first
breakage time $W(\tau)$, the mean
extension of the bonds with respect to the consecutive circle number in the
membrane, as well as other quantities of interest.

Since in the problem of thermal degradation there is no external force acting on
the membrane edges, a well-defined activation barrier for a bond scission is
actually missing, in contrast to the case of applied tensile force. Therefore,
a definition of an unambiguous criterion for bond breakage is not self-evident.
Moreover, depending on the degree of stretching, bonds may break and then
recombine again. Therefore, in our numeric experiments we use a sufficiently
large value for critical extension of the bonds, $r_h=5r_{\text{min}}$, which is
defined as a threshold to a broken state of the bond. This convention is based
on our checks that the probability for recombination (self-healing) of bonds,
stretched beyond $r_h$, is sufficiently small, $< 10^{-5}$ .

Also we examine the course of the degradation kinetics: at periodic intervals
in separate
simulations runs
we
analyze the size distribution of fragments (clusters) of the initial membrane
and establish the time-dependent probability distribution function of fragment
sizes, $P(n,t)$, as time elapses after the onset of the thermal degradation
process. This also yields the time evolution of the mean fragment size, $N(t) =
\int_0^{\infty} n(t) P(n,t) dn $. We perform the statistical averaging of
fragment sizes by developing an appropriate for the system fast cluster counting
algorithm.


The Molecular Dynamics calculations were carried out using a cluster counting
algorithm based on an {\em ad hoc} implementation of the Hoshen-Kopelman
program \cite{Hoshen_Kopelman}.
In a subsequent paper of Al-Futaisi and Patzek \cite{Patzek}, the
Hoshen-Kopelman
algorithm for cluster labeling has been extended to non-lattice environments
where network elements (sites or bonds) are placed at random points in space.
Following  Al-Futaisi and Patzek \cite{Patzek}, we developed our
simplified version of this algorithm which concerns only clusters of nodes in a
network with honeycomb structure (which is our model shown in
Fig.~\ref{fig_EquilMembrane}).
In our implementation  we assume that all nodes
 in the network are occupied, but some links can be broken when we
study thermal degradation process of the network. The network information is
stored in two arrays. The first one is related to the connectivity of the nodes
and the second to the state of links (intact or broken link)

\section{RESULTS and DISCUSSION}\label{sec_MD_results}
\subsection{Bond scission}
\label{subsec_tau}

\begin{figure}[!h]
\begin{center}
\vspace{-0.5cm}
\includegraphics[scale=0.20]{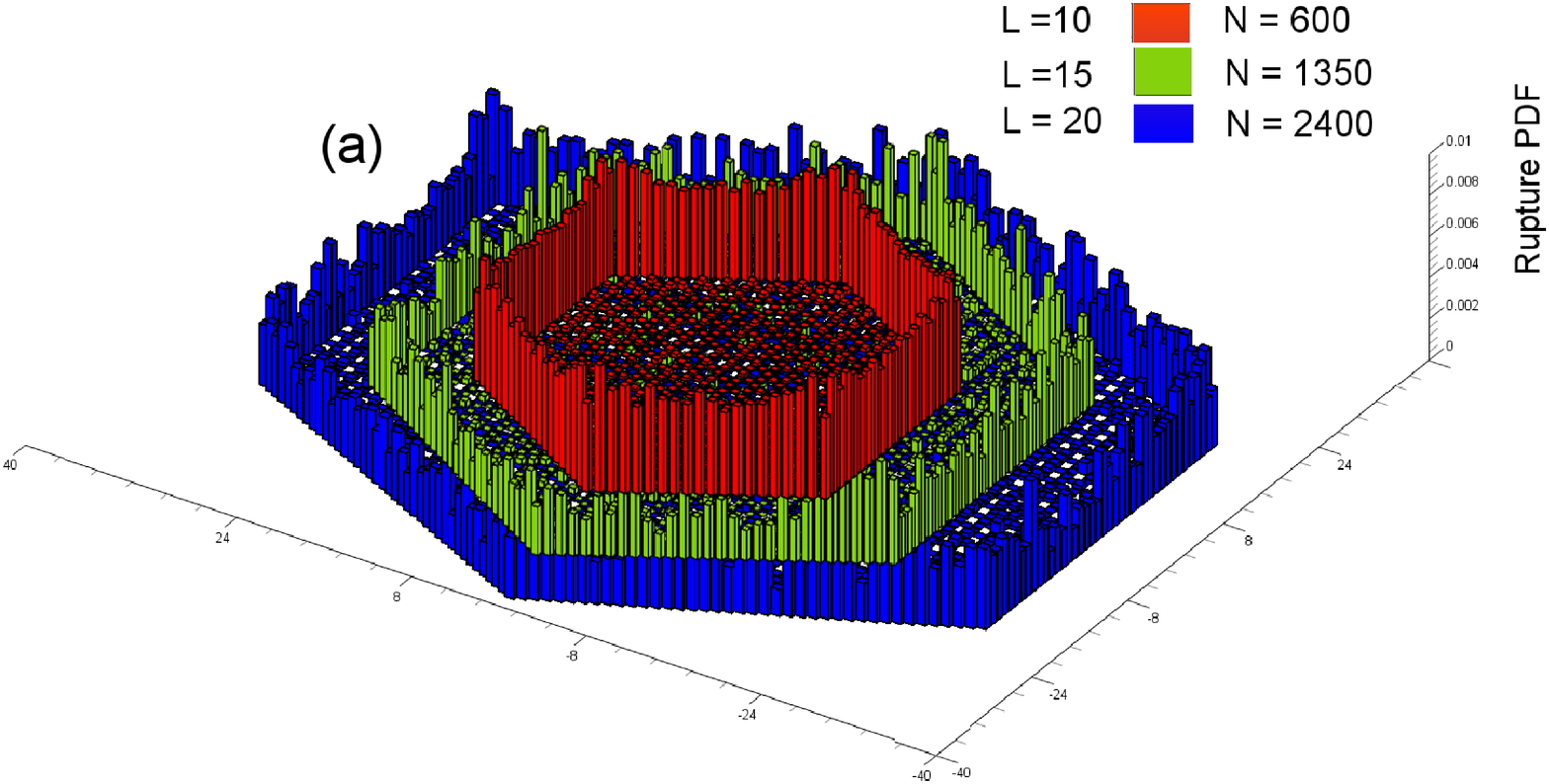}
\hspace{0.5cm}
\includegraphics[scale=0.14]{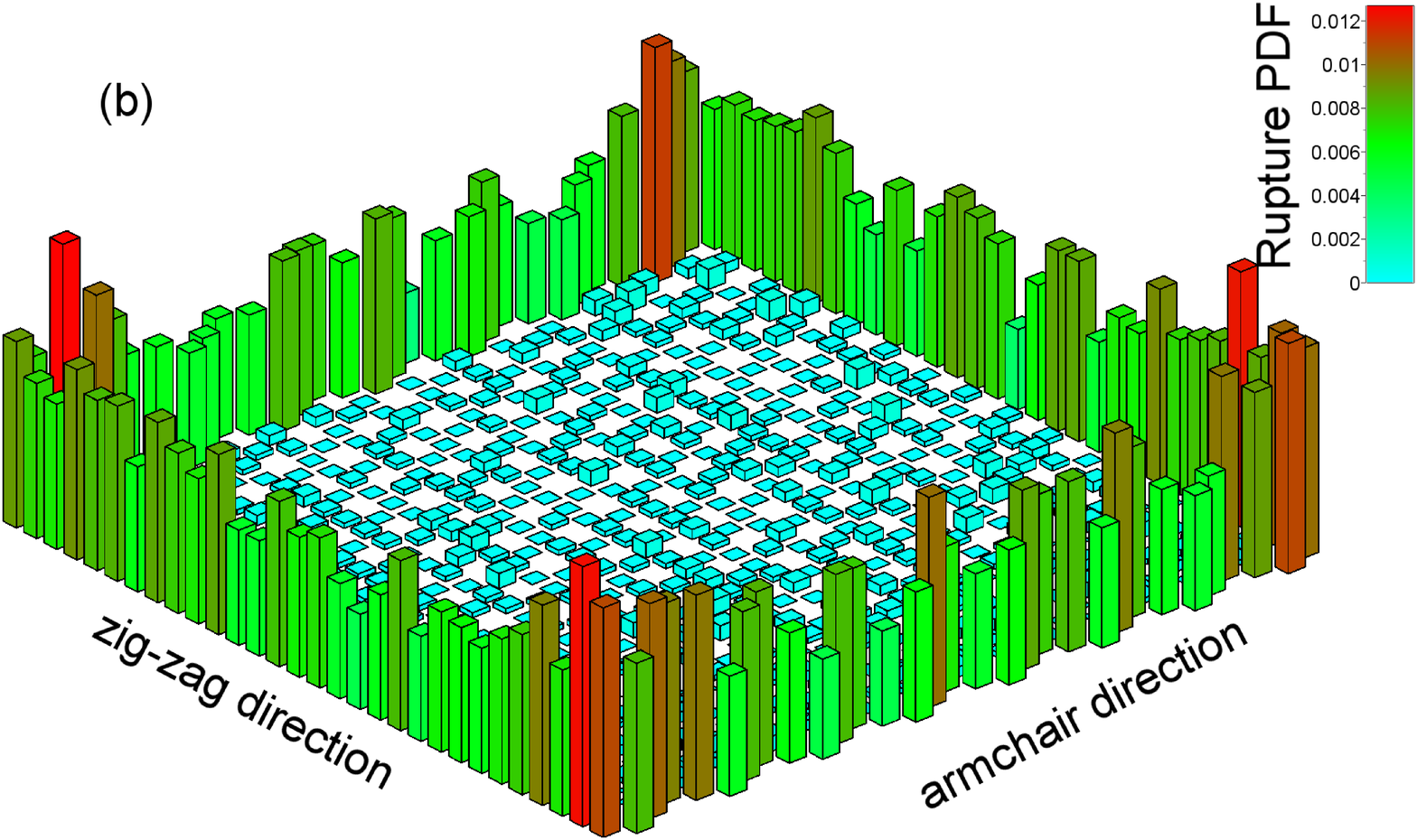}
\\
\phantom{.}
\hspace{-0.5cm}
\includegraphics[scale=0.2]{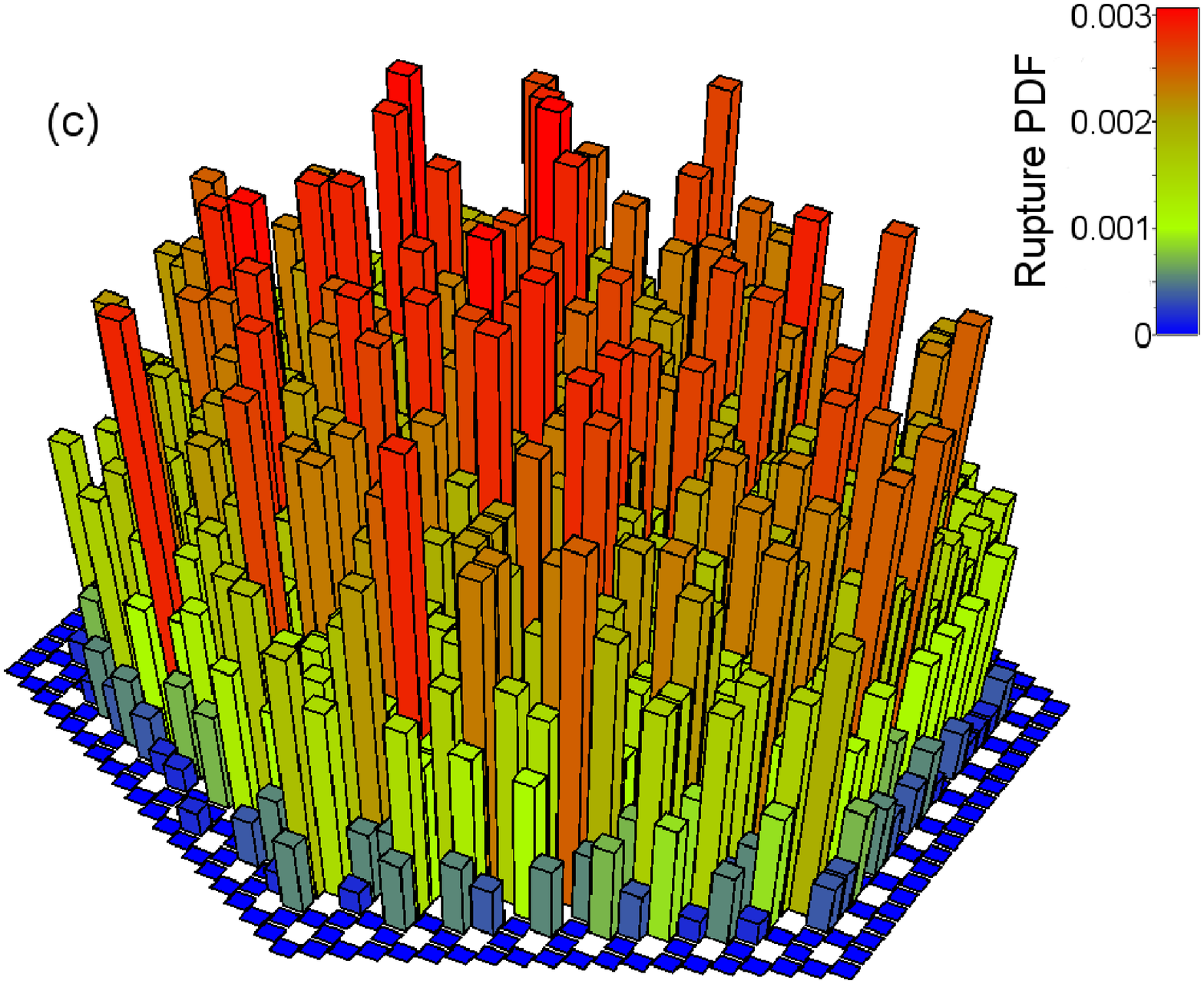}\\
\phantom{.}
\hspace{-0.5cm}
\includegraphics[scale=0.21]{hist_circ.eps}
\caption{(a) Rupture probability histograms for thermally induced scission
events
in flexible honeycomb flake of different linear size $L$  as shown in the
legend. (b) Rupture histogram for a ribbon-like square honeycomb membrane with
$496$ nodes. 
(c) Rupture histogram for a membrane flake with beads tethered at the rim,
$L=10$.
Parameters of the heat bath are $T=0.1$ and $\gamma=0.25$.
(d)
Probability distribution of breakage events as a function of consecutive circle
number for a membrane flake with $N=600$ and two different friction
coefficients $\gamma = 0.25$ and $10.0$. Here $T=0.1$.  The inset shows
estimated values of Lyapunov exponents $\lambda$ vs $T$ for beads located 
in the rim/bulk of membrane as indicate. 
Here $N=5400$, $\gamma=0.25$.
 \label{fig_histo} }
\end{center}
\end{figure}

In Fig.~\ref{fig_histo}a we show the distribution of bond scission rates among
all bonds of the honeycomb membrane for flakes (with zig-zag pattern on the
periphery) of several sizes $L = 10,\; 15,\; 20$. Somewhat surprisingly, one
finds the  overwhelming fraction of bond breaking occurs at the outer-most
rim of the membrane where monomers are bound by only two bonds to the rest of
the sheet. We have also sampled the rupture histograms for ribbon-like square
membranes, Fig.~\ref{fig_histo}b. Interestingly, we observe no difference
between the rupture rates of zig-zag and armchair edges whereas the bonds in the
four corners of such membrane expectedly break more frequently. The difference
in the relative stability of the bonds becomes clearly evident in
Fig.~\ref{fig_histo}d where the frequency of periphery bonds appears nearly
two orders of magnitude larger when compared to bonds in the 'bulk' of the
membrane where each monomer (node) is connected by three bonds to its neighbors.
One may therefore conclude that a moderate increase in the coordination number
of the nodes (by only 33\% regarding the maximum coordination of a node) leads
to a major stabilization of the supporting bonds and much stronger resistance to
fracture. Our additional simulation in the strongly damped regime for
$\gamma=10$ indicates no qualitative changes compared to $\gamma=0.25$ except
an absolute overall increase of the rupture times which is natural for a more
viscous environment.

Note that the question of where and which bonds predominantly break is by no
means trivial. For example, in the case of linear polymer chain thermal
decomposition the rate of bond rupture is least at both chain ends although
the end monomers, in contrast to those inside the chain, are bound by a single
bond only \cite{Paturej}. We demonstrate in Fig.~\ref{fig_histo}c that this
interesting feature holds also for the honeycomb membrane flake, provided the
rim is clamped and left immobile during the simulation. Evidently, the highest
 frequency of bond scissions grows towards the membrane center. Similar to the
case of polymer chains with fixed ends \cite{Paturej, Lee}. 

In order to provide deeper insight into the mechanism of temperature-induced
bond breaking, in the inset to Fig.~\ref{fig_histo}d we present the temperature
variation of Lyapunov's exponent $\lambda$ for membrane nodes located in the
bulk and in the rim of the sheet. Evidently, beyond a cross-over temperature
$T\approx 0.05$ one observes a significant growth of $\lambda_{\mbox{\tiny
rim}}$ as compared to $\lambda_{\mbox{\tiny bulk}}$. This indicates that the
trajectories of nodes at the membrane periphery attain much faster chaotic
features at higher temperature than those of the bulk nodes. Moreover, we should
note that beads in the vortices have values of $\lambda$ which exceed those in
the rim by about $5\%$. Therefore this analysis of trajectory stability at
characteristic locations in the membrane clearly demonstrate that bond rupture
is induced by intermittent motion of the respective nodes. 

\begin{figure}[!h]
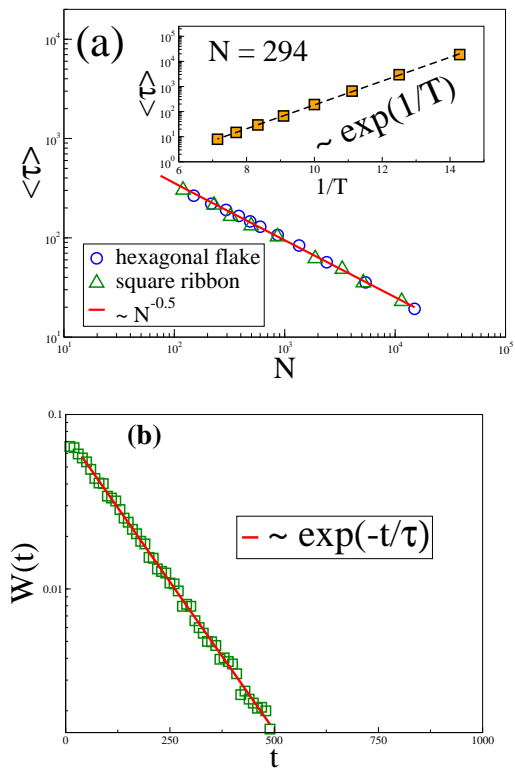

\begin{center}
\includegraphics[scale=0.25]{tauvsNthermo.eps}
\\
\phantom{.}
\hspace{-0.5cm}
\includegraphics[scale=0.25]{Wtau.eps}
\caption{(a) Mean first breakage time $\langle \tau \rangle$ vs. number of beads
$N$  for two different membrane shapes: a hexagonal flake and a square ribbon.
Solid line represents a fit by power law with an exponent $\beta=0.5$ in both
cases. The inset shows the variation of $\langle \tau \rangle$ with inverse
temperature $1/T$ for a flake membrane with $N=294$ particles. The fitting line
yields an Arrhenian relationship, $\langle \tau \rangle \propto \exp(\Delta
E_b/k_BT)$ with dissociation energy barrier $\Delta E_b \approx 0.95$. (b)
Breakage time probability distributions $W(t)$ for thermally induced breakage of
a flake membrane with $N=600$ particles at $T=0.1$. Symbols denote result of the
simulation and full line stands for a fitting function $W(t) \propto \exp(-t /
\tau)$ with $\tau=128.2$. \label{fig_MFBT}} \end{center} \end{figure}

The variation of the MFBT $\tau$ of a bond with membrane size $N$ during
thermolysis for both hexagonal and square shapes of the 2D sheet is displayed in
Fig.~\ref{fig_MFBT}a. Evidently, one observes for $\tau$ a well pronounced power
law behavior, $\tau \propto N^{-\beta}$ with an exponent $\beta \approx 0.50 \pm
0.03$. It turns out that  the scaling exponent $\beta$ remains insensitive to
changes in the geometric shape of the membrane sheet. This value of $\beta$ might
appear somewhat surprisingly to deviate from the expected exponent of unity,
given that in the absence of external force all bonds are supposed to break
completely at random so that the total probability for a bond scission (i.e.,
the chance that {\em any} bond might break within a time interval) is additive
and should be, therefore, proportional to the total number of available bonds,
$N_{bonds} = (3N - 6L)/2$. As suggested by Fig.~\ref{fig_histo}, however,
predominantly only periphery bonds are found to undergo scission during thermal
degradation. The number of periphery bonds goes roughly as $\propto \sqrt{N}$
which agrees well with the observed value $\beta \approx 0.5$ and provides a
plausible interpretation of the simulation result, Fig.~\ref{fig_MFBT}a. From
the inset in Fig.~\ref{fig_MFBT}a one may verify that the bond scission displays
an Arrhenian dependence on inverse temperature, $\tau \propto \exp(\Delta E_b /
k_BT)$, with a slope $\Delta E_b \approx 1$. This slope suggests a
dissociation energy $\Delta E_b$ of the order of the potential well depth of the
Morse interaction, Eq.~(\ref{eq_U_MORSE}) where $\epsilon_M = 1.0$.
In our model we deal typically with
$E_b/(k_BT)\approx 10$ which at $300$~K and typical bond length
$r_{\mbox{\tiny min}}\approx 0.144$~nm,
corresponds to ultimate tensile stress $\sim 0.6$~GPa. This is a reasonable
value for our membrane which is considerably softer than graphene with
$\sim100$~GPa \cite{Aluru} and is ranged between typical values for rubber
materials $~0.01$--$0.1$~GPa.

The probability distribution function (PDF) of MFBT $W(t)$, i.e., the PDF of
the time interval before the first breakage event in the membrane takes place,
is shown in Fig.~\ref{fig_MFBT}b for $T = 0.1$. Evidently, one has $W(t) \propto
\exp(- t / \tau)$ with a sharp maximum close to $t \approx 0$. At even lower
temperature one might expect this maximum to become more pronounced, suggesting
 thus a
Poisson distribution for the $W(t)$. As far as $\tau$ tends to grow
exponentially fast with decreasing $T$ - see inset in Fig.~\ref{fig_MFBT}a,
collecting statistics in this temperature range becomes difficult.

\begin{figure}[!h]
\begin{center}
\includegraphics[scale=0.3]{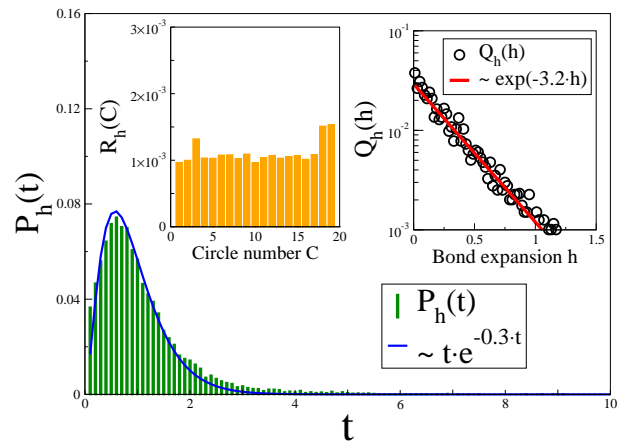}
\caption{Probability distribution $P_h(t)$ of times (impulses),
and $Q_h(h)$ of  bond overstretching lengths $h$ (circles, right panel of inset)
before
a recombination event in a membrane with $N=600$, $T=0.1$, $\gamma=0.25$ takes
places. The recombination times probability distribution $P_h(t)$ is fitted by
Poisson distribution (blue line). The probability for a bond stretching a
distance $h$ beyond $r_{\mbox{\tiny inflex}}$ is described by $Q_h(h)$
indicating an
exponential decay - red
line.
The left inset shows healing probability $R_h$ vs.~consecutive {\em circle}
number, demonstrating which part of the membrane undergoes healing most
frequently.  \label{fig_heal}}
\end{center}
\end{figure}

We also analyze the fracture of bonds with respect to their possible
recombination.
The procedure of sampling bond recombination events is the following. Once one
of the bonds in the membrane is stretched to the position of the Morse potential
inflexion point  $r_{\mbox{\tiny inflex}}\approx 2.69$ (corresponding to maximal
tensile
strength of a given bond), we start to monitor its further expansion for $10^4$
integration steps and record its maximal expansion $h=r-r_{\mbox{\tiny
inflex}}$. If the
bond
subsequently shrinks to  $r<r_{\mbox{\tiny inflex}}$, this is counted as
recombination event.
Simultaneously we record the time $t$ needed to go back below the expansion
$r_{\mbox{\tiny inflex}}$. We then compute distributions  $Q_h(h)$ of bond
expansions beyond
$r_{\mbox{\tiny inflex}}$ and recombination times $P_h(t)$. Thus $Q_h(h)$ yields
the
probability of bond overstretching to distance $h=r-r_{\mbox{\tiny inflex}}$
during the
recombination event. As indicated by the distribution $P_h(t)$ shown in
Fig.~\ref{fig_heal}, the interval $20$ t.u.~exceeds more than five times
the maximal life of recombination which guarantees that all such events are
properly taken into account. Generally we find that a recombination of bonds
takes place rather seldom (roughly once per simulation run of average length 130
t.u., as given by the MFBT).

It has been mentioned in Sec.~\ref{sec:model} that in the absence of
external force acting on the membrane the criterion for a bond to be considered
broken is not unambiguous. Adopting a rather large critical bond extension of
$r_h = 5 r_{\mbox{\tiny min}}$ practically rules out the probability of of
subsequent
recombination of such bonds, as indicated by $Q_h(h) \propto \exp(- 3.2 h)$ in
\ref{fig_heal} (right inset).
 Indeed, if one adopts $r_{\mbox{\tiny inflex}}$ as a threshold for rupture than
$Q_h(h)$
suggests that bond expansions larger than
$r_{\mbox{\tiny inflex}}+1.1\approx 3.79$
practically never happen.
 The
PDF $P(t)$ of times elapsed before self-healing is also shown in
Fig.~\ref{fig_heal} where it appears as a Poisson distribution, $P(t) \approx
t\exp(-0.3 t)$. Not surprisingly, most of the recombination occur
in the periphery of the membrane - left inset in Fig.~\ref{fig_heal}.
\begin{figure}[!h]
\begin{center}
\includegraphics[scale=0.3]{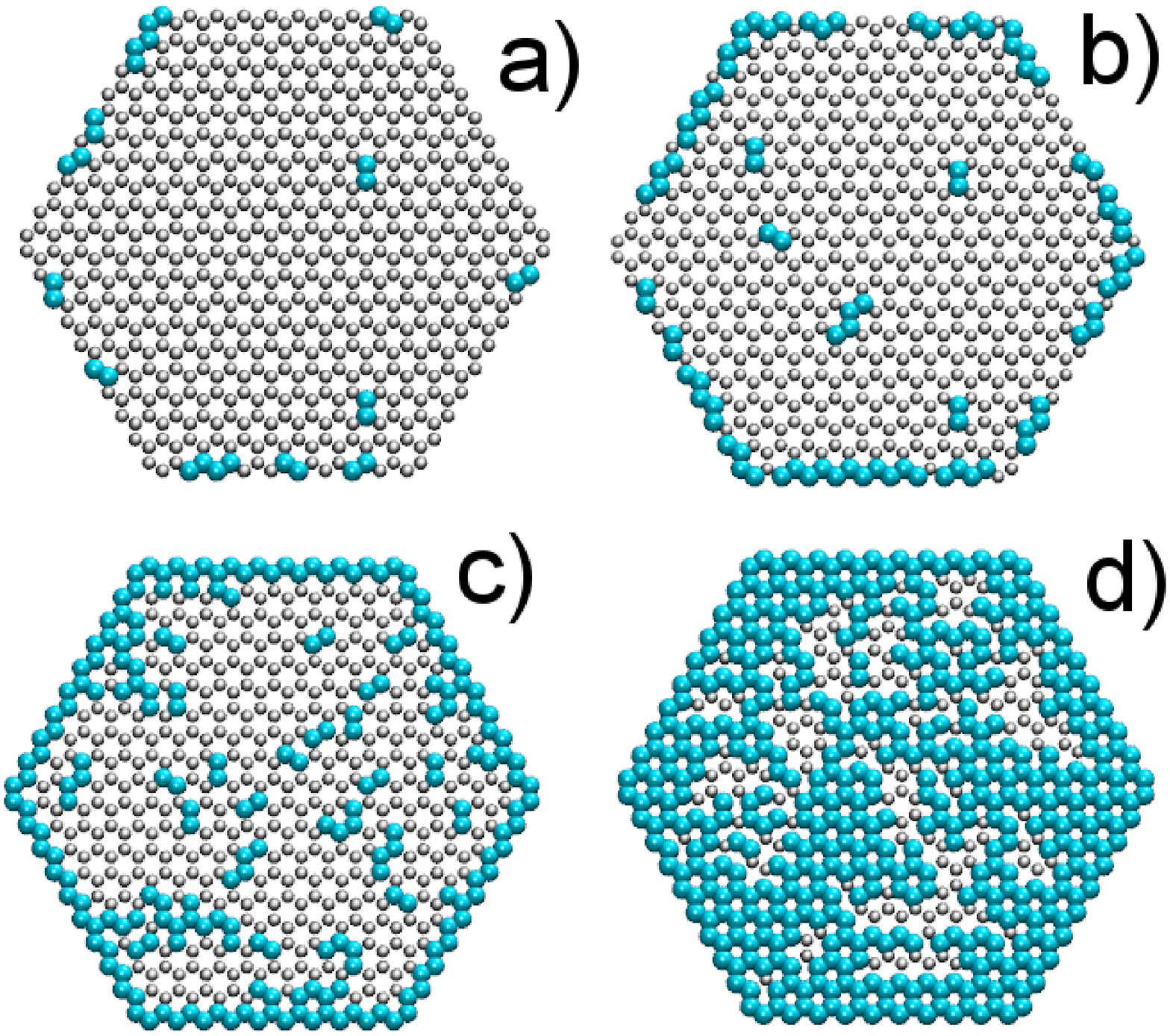}
\hspace{0.5cm}
\includegraphics[scale=0.191]{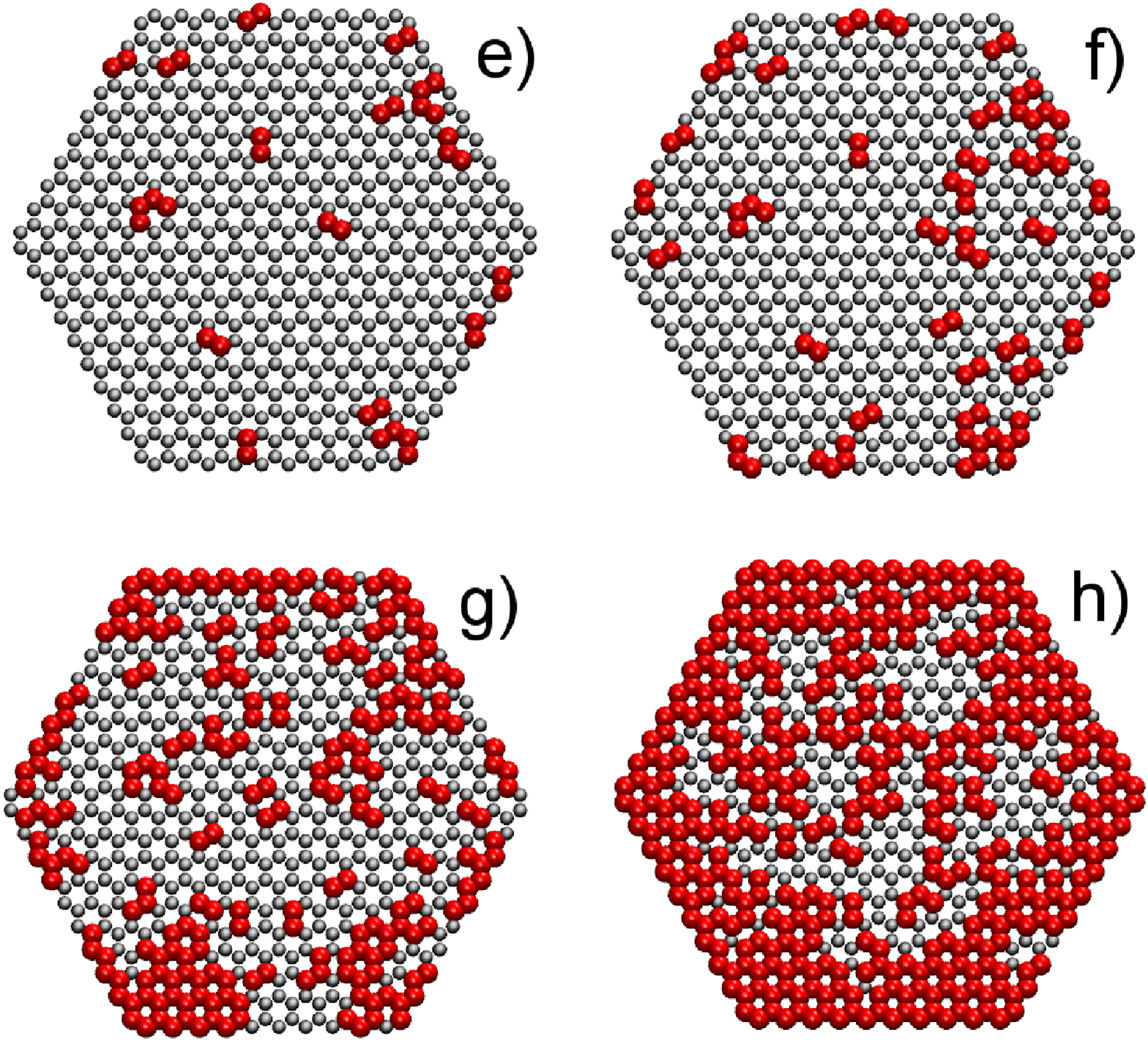}
\caption{Thermal breakage of bonds in a membrane made of $N=600$ particles at
different time moments: a) $t=10^2$, b) $7\cdot 10^2$ c) $5\cdot 10^4$ d)
$3\cdot 10^5$ and e) $50$, f) $100$, g) $250$, h) $500$. Broken bonds are marked
by blue ($T=0.1$) or red ($T=0.15$) color, depending on the temperature of a
heat bath, while gray color corresponds to intact bonds. \label{fig_cracks} }
\end{center}
\end{figure}

In Fig.~\ref{fig_cracks} we show the distribution of broken bonds at different
times after the onset of thermal degradation.
We would like to note that the regular hexagonal flakes shown in
Fig.~\ref{fig_cracks}
serve only to indicate schematically the positions of   both broken
and still intact bonds
and by no means represent the actual conformation of the membrane.
 Two different temperatures, $T =
0.10$ and $T = 0.15$ are studied. One can readily verify from
Fig.~\ref{fig_cracks}a-d that at $T = 0.10$ the degradation process starts from
the rim of the network sheet and then proceeds inwards, as noted
earlier by Meakin \cite{Meakin}. In contrast, at $50\%$ higher temperature,
i.e.,
at $T = 0.15$, bonds break everywhere in the network sheet - cf.
Fig.~\ref{fig_cracks}b,~f and Fig.~\ref{fig_cracks}c,~g. Such difference in the
bond scission mechanism at different temperatures has been observed
\cite{Meakin} before. It appears that at  $T=0.1$ the membrane periphery
undergoes stronger oscillations than the membrane bulk 
which lead to bond scission at the rim while in the inner part
of the sheet the monomers mutually block each other and the bonds remain
largely intact. This explanation is supported by the measured values of Lyapunov
exponents (see the inset of Fig.~\ref{fig_histo} and discussion in
Sec.~\ref{sec_MD_results} A).
On the other hand a 'hot' i.e.~at $T=0.15$ membrane undergoes
much stronger agitation as a whole
and as a result bonds break all over the network sheet. Moreover, at lower $T$
when the thermal agitation of the membrane is weaker, one may correlate the
probability of bond scission with the average strain in the bonds -
\begin{figure}[!h]
\begin{center}
\includegraphics[scale=0.25]{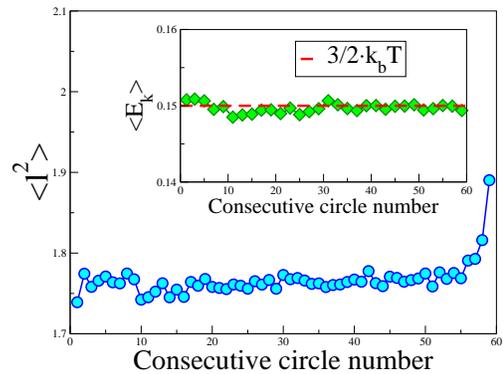}
\caption{ Variation of mean squared bond length $\langle l^2\rangle$
vs.~distance from membrane center (consecutive circle number) for thermolysis
of membrane composed of $N=5400$ beads. The inset displays mean kinetic energy
$\langle E_{k}\rangle$ (per monomer) as a function of circle number. Red dashed
line represents the level of energy which corresponds to equipartition theorem.
Parameters of thermostat are $T=0.1$ and $\gamma=0.25$. \label{fig_bondlength}}
\end{center}
\end{figure}
Fig.~\ref{fig_bondlength}. Evidently, the average squared bond length $\langle
l^2 \rangle$ increases steadily and becomes nearly $8\%$ larger for the bonds
sitting on the last ring of the membrane whereas no difference in the mean
kinetic energy between peripheral and bulk nodes is detected -
Fig.~\ref{fig_bondlength}. Note that the kinetic energy (that is, the
temperature $T$) is thereby uniformly distributed along the sheet - see inset
in Fig.~\ref{fig_bondlength}.

\subsection{Temporal evolution of the fragmentation process}
\label{subsec_fragment}

\begin{figure}[ht]
\begin{center}
\includegraphics[width=0.36\textwidth]{fragment.eps}
\hspace{0.50cm}
\includegraphics[totalheight = 0.6\textwidth, origin = br,
angle=270]{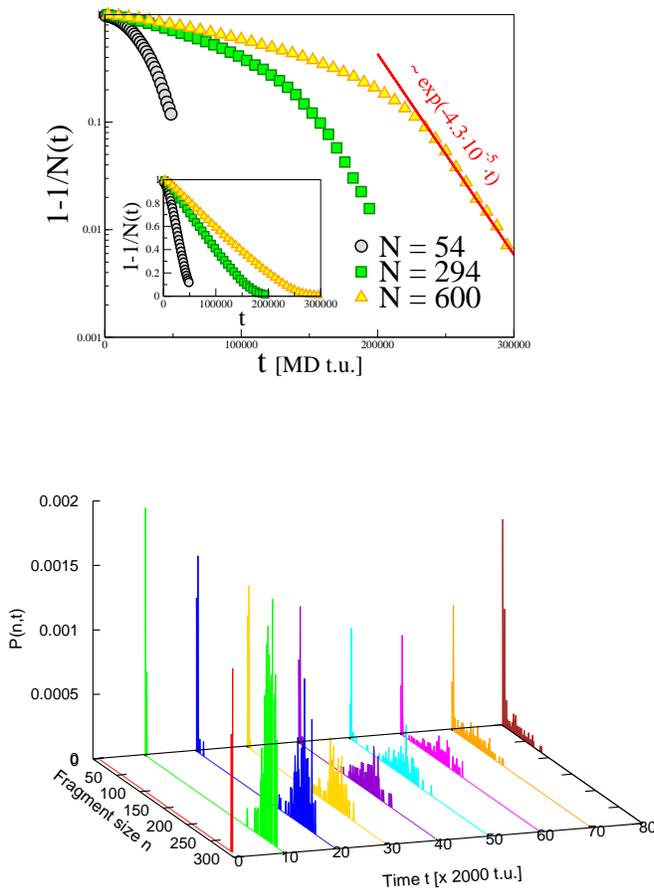} \caption{(a) Semi-log plot of
time variation of
mean fragment size $N(t)$ for membranes made of different size as indicated.
Symbols represent simulation results whereas a red line stands for the fitting
function $1-1/N(t) \propto \exp{(-kt)}$. The kinetic constant $k = 4.3 \time
10^{-5} s^{-1}$.  In the inset the same is shown in normal coordinates. (b)
Probability distribution of fragment sizes $P(n,t)$ at different times $t$ (in
MD time units) after beginning of the thermal degradation process for a membrane
with $N=294$. Parameters of a heat bath are $T=0.12$ and $\gamma=0.25$.
\label{fig_PDF}}
\end{center}
\end{figure}

After the onset of the thermal decomposition process the membrane flake
disintegrates into smaller clusters (fragments) of size $n$ whose mean size
(or average
molecular weight) $N(t)$ decreases steadily with time. $N(t)$ is easily
accessible experimentally, therefore we give in Fig.~\ref{fig_PDF}a the course
of its temporal evolution, observed in our computer experiment. Using an {\em ad
hoc} cluster counting program in the course of the MD-simulation we sample the
probability distribution of fragment sizes, $P(n,t)$, so that the first moment
$N(t) = \int n(t) P(n,t) dn$ gives the cluster mean size $N(t)$. Thus, for a
given time moment $t$ we average data over more than $10^3$ independent runs,
each starting from a different initial conformation of the honeycomb membrane.
In Fig.~\ref{fig_PDF}b we show the time variation of the ensuing PDF $P(n,t)$
whereby the system is seen to start with a single sharp peak at $t = 0$ when the
membrane is still intact. As time goes by, $P(n,t)$ becomes bimodal, the maximum
of the distribution is seen to shift to smaller values of cluster size whereas
an accumulation of fragments of size $1$ or $2$ is observed to contribute to a
second peak at
$n \approx 1$. Eventually, as $t \rightarrow \infty$, the PDF $P(n,t)$ settles
to a shape with a single sharp peak (a $\delta$-function) at $n \approx 1$ (not
shown here).

One can readily see from Fig.~\ref{fig_PDF}a that the quantity $1 - N^{-1}(t)$
does not immediately follow a straight line of decay when plotted in
semi-logarithmic coordinates, rather, such a decay is observed after an initial
period of slower decline. This effect is due to averaging over many realizations
of the fragmentation process. In each run the degradation of bonds starts
earlier or later at a particular time $\tau$ (the Mean First Breakage Time) that
is
distributed according to $W(\tau)$ - cf.~Fig.~\ref{fig_MFBT}b. As a result a
clear cut exponential course of $1 - N^{-1}(t)$ is only observed in the late
stages of fragmentation. Such behavior is found independently of the membrane
size - Fig.~\ref{fig_PDF}a.

In addition, one could expect that the fragmentation process is not governed by
a single rate constant in a presumably $1^{st}-$order chemical reaction even
though the bonds that undergo rupture are chemically identical. Therefore, from
the temporal mean cluster size behavior, presented in Fig.~\ref{fig_PDF}a, one
may conclude that even in the case of a homogeneous membrane the thermal
degradation process is more adequately described by several reaction constants
which govern the dissociation of different groups of bonds. In the next section
we suggest a simple model of reaction kinetics which takes into account this
conjecture.

\subsection{Reaction kinetics}
\label{sec_diff_eq}

One may try to reproduce the bond scission kinetics during in the course of
thermal fragmentation by a set of several $1^{st}$-order chemical reactions. The
nodes of the membrane can be subdivided into several groups, depending on the
number of intact bonds that connect them to neighboring nodes. In the case of a
honeycomb membrane one may distinguish four such groups, and denote the
instantaneous number of such nodes (monomers) by $n_0,\;n_1,\;n_2$ and $n_3$
whenever $0,\;1,\;2,$ or $3$ intact bonds exist around such a node. If
self-recombination of bonds is ignored, which is reasonable in view of large
value of the threshold, and simultaneous scission of two and more bonds is
disregarded as hardly probable, one can write down a strongly simplified set of
$1^{st}$-order kinetic differential equations (DE) that describes the evolution
of $n_0(t),\; n_1(t),\; n_2(t)$ and $n_3(t)$ with time:
\begin{eqnarray}\label{DE}
 \dot{n}_1(t) &=& -k_1 n_1(t) + k_2 n_2(t)\\ \nonumber
 \dot{n}_2(t) &=& -k_2 n_2(t) + k_3 n_3(t)\\ \nonumber
 \dot{n}_3(t) &=& -k_3 n_3(t)
\end{eqnarray}
Thus, for example, the number of double-bonded nodes decreases as one of the two
bonds breaks, however, if a bond breaks around a node with triple coordination
this would increase the population of double-bonded nodes. Note, that the
total number of nodes of all kinds remains thereby conserved,
\begin{equation} \label{conserve}
  n_0(t) + n_1(t) + n_2(t) + n_3(t) = N = 6 L^2.
\end{equation}

If the degradation process starts with an intact membrane conformation at $t =
0$ (no broken bonds exist), one can fix the initial conditions as $n_0(0) = 0,\;
n_1(0) = 0, n_2(0) = 6L$ and $n_3(0) = N - n_2(0) = 6L (L-1)$. Thus, initially
one has only  $n_2(0) = 6L$ double-bonded nodes at the membrane periphery along
with $n_3(0) = 6L (L-1)$ triple-bonded nodes in the bulk of the flake. Then,
as the decomposition process develops, nodes from a given class $r = 1,\;2,\;3$
will transform into a lower class $r = 0,\;1,\;2$ ones (we neglect hereby the
simultaneous scission of more than one bond of a node as highly improbable
event). Eventually, at $t \rightarrow \infty$ the fragmentation process ends and
one expects $n_1(\infty) = n_2(\infty) = n_3(\infty) = 0$ and $n_0(\infty) = N$.

One may solve analytically the system of $1^{st}$-order DE Eqs.~(\ref{DE}) to a
set
of functions $n_i(t)$ with $0 \le  i \le 3$:
\begin{widetext}
\begin{eqnarray}
  n_0(t) &=& 6 L^2 - n_1(t) - n_2(t) - n_3(t) \nonumber \\
  n_1(t) &=& \frac{6 L}{(k_1-k_2)(k_1-k_3)(k_2-k_3)}
             \left[ k_1 (k_2-k_3) \left(e^{-k_3 t} - e^{-k_1 t}\right) +
              k_2 k_3 \left(e^{-k_3 t} - e^{-k_2 t}\right) \right. \cr
         &+&
      \left. k_1 k_3 L \left(e^{-k_3 t} - e^{-k_2 t}\right) -
      k_2 k_3 L \left( e^{-k_3 t} - e^{-k_1 t}\right)
      + k_3^2 L \left(e^{-k_2 t} - e^{-k_1 t}\right) \right] \nonumber \\
  n_2(t) &=& 6 L \frac{\left(k_2 e^{-k_2 t} - k_3 e^{-k_3 t}
- k_3 L e^{-k_2 t} + k_3 L e^{-k_3 t} \right)}{k_2 -k_3} \\
  n_3(t) &=& 6 L (L-1) e^{-k_3 t} \nonumber
\label{DE_sol}
\end{eqnarray}
\end{widetext}
where the rate constants $k_1,\;k_2$ and $k_3$ are still to be determined, for
example, by comparison with simulation data.  As one may readily verify from
 Fig.~\ref{fig_histo}, our simulation data suggest that a bond to a
triple-bonded node at $T = 0.12$  is much more stable (by about two orders of
magnitude) than a bond at the flake periphery where each node is connected by
only two bonds to the network. Thus the system of Eqs.~(6) can be
tested for a set of reaction constants $k_1 \gg k_2 >  k_3$ directly by means of
our Molecular Dynamics computer experiment.

\begin{figure}[ht]
\begin{center}
\vspace{0.5cm}
\includegraphics[width=0.36\textwidth]{N54_new.eps}
\hspace{0.5cm}
\includegraphics[totalheight = 0.4\textwidth, origin = br, angle =
270]{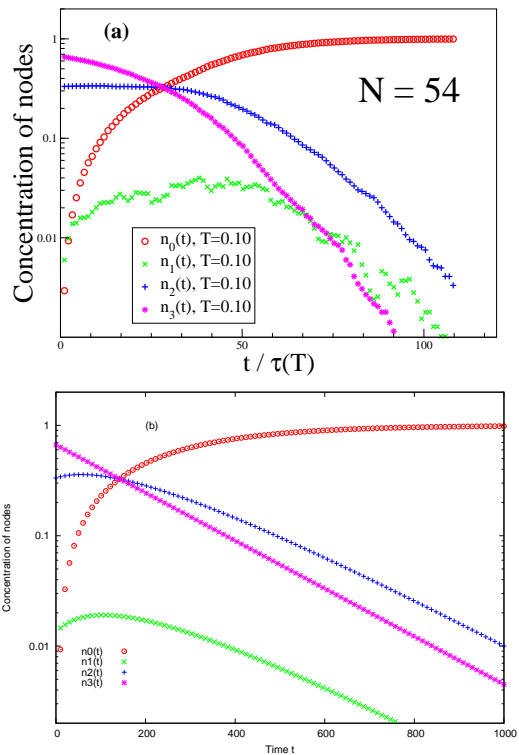}
\caption{(a) Variation of the number of nodes $n_m(t)$ with $m = 0$--$3$
intact
bonds with elapsed dimensionless time $t / \tau(T)$ for a membrane with
$N = 54$ and  $T=0.10$. Here $\tau(T)$
denotes the characteristic time of degradation at the respective temperature
$T$. The $n_1(t)$ (monomers bound by a single bond) are shown by shaded area.
(b) The same as in (a) for the same membrane size $N = 54$ according to the
analytic result Eqs.~(6). The values of the kinetics rate constants
are $k_1 = 20.0,\; k_2 = 0.007,\; k_3 = 0.005$.
\label{fig_nodes}}
\end{center}
\end{figure}

Indeed, one finds by comparing the simulation result, Fig.~\ref{fig_nodes}a for
a membrane of size $N = 54$ at $T = 0.10$, and the analytical solution,
Eqs.~(6), Fig.~\ref{fig_nodes}b, that the observed kinetics agree
qualitatively, provided one allows for the absence of fluctuation  in
Eqs.~(4) (i.e., for the disarray and averaging of the MFBT
$\tau$ in the simulation data). We should like to point out here that the
values of the rate constants $k_1,\; k_2,\; k_3$ are {\em not} best fit values.
Because of the
additional effect of averaging discussed in Sec.~\ref{subsec_fragment} and
shown in Fig.~\ref{fig_PDF}, a straight fitting
procedure with three parameters $k_1,k_2,k_3$ would be both inefficient and
hardly successful. Therefore we tried different combinations of values for
$k_1,k_2,k_3$ subject to the condition that $k_1\gg k_2>k_3$.
Even though
the general qualitative shapes turn out to be rather sensitive to the values of
$k_1,k_2,k_3$ (being capable of reproducing, e.g., the existence of a common
intersection point of $n_0(t), n_2(t) $ and $n_3(t)$ - see Fig.~\ref{fig_nodes}a
- for particular choice of parameters), one finds thus easily a combination
which qualitatively matches
well the simulation data shown
in Fig.~\ref{fig_nodes}a.

One may conclude therefore that the simplified set of $1^{st}-$order DE
Eqs.~(\ref{DE}) captures qualitatively the main features of the fragmentation
kinetics and the principal mechanism at work is a combination of few
$1^{st}-$order chemical reactions of bond scission. Nonetheless, it is
conceivable to expect that for a full quantitative description of the thermal
degradation process the set of kinetic equations, Eqs.~(4), should be
extended by few additional reactions: $n_1(t) \rightarrow n_2(t),\; n_2(t)
\rightarrow n_3(t)$ that may in principle also take place (with respective rate
constants). One can then still derive an analytical solution of the extended set
of DE that describes the full kinetics of fragmentation and try to fit the
ensuing rate constants to the simulation data.  In view of the growing number
of fit parameters, however, a detailed analysis of such system is beyond the
scope of the present work and should be left for future work.

\section{CONCLUSION}\label{sec_summary}

In the present investigation we use Langevin Molecular Dynamics simulation and
also solve a set of $1^{st}-$order kinetic DE so as to model the
process of thermal destruction of a polymerized membrane sheet with honeycomb
structure.
Results of our work can be treated as generally applicable due to the fact that
a two-dimensional regular lattice can only be created in few ways: honeycomb,
triangular and square lattice with second neighbor interactions because of the
zero-shear modulus (apart from exotic cases like quasi-crystalline, kagome,
etc. lattices of little relevance).
The differences in lattice coordination among the first three periodic networks
induce only quantitative renormalization of the Young modulus. However they
don't change the overall elastic behavior. 
 Our findings regarding the most salient features of thermolysis in an
 elastic brittle honeycomb network sheet subject to sufficiently high
temperature can be
summarized as follows:
\begin{itemize}
 \item The probability of bond scission is highest at the periphery of the
membrane sheet where nodes are connected by two bonds only. At higher
temperature, however, the whole sheet undergoes fragmentation whereby also bonds
in the bulk rupture.
 \item The mean time $\tau$ until a bond undergoes scission
event declines with the number of nodes $N$ (with membrane size) by a power law
as
$\tau \propto N^{-0.5}$ independently of the geometric
shape of membrane sheet.
 The times of bond scission are exponentially
distributed, $W(t) \propto \exp(- t / \langle \tau \rangle)$.
 \item Bond recombination (self-healing) occurs seldom during thermal
degradation, and the measured recombination times follow Poisson distribution
whereas an extra-stretching of bonds (beyond the point of maximal tensile
strength)
 before a recombination takes place is exponentially improbable.
 \item the fragmentation kinetics is determined by $1^{st}$-order reactions
between network nodes with different number of intact bonds and follows simple
exponential decay at late times.
 \item A set of $1^{st}$-order kinetic differential equations, describing the
process of fragmentation, can be established and solved analytically. One finds
results in qualitative agreement with those from computer experiment providing
thereby a deeper insight into the mechanism of thermal degradation of
two-dimensional honeycomb networks.
\end{itemize}

In view of the presented results, it should be clear that more work is needed
(e.g., regarding the effects of randomness in brittle networks)
until a full understanding of the process of thermal degradation in polymerized
$2D$-membranes is reached.

\section{Acknowledgments}

J.P.~would like to thank the Institute of Physical Chemistry at the Bulgarian
Academy of Sciences for hospitality during his stay.
A.~M.~gratefully acknowledges support by the Max-Planck-Institute for Polymer
Research during the time of this investigation. This study has been supported
by the Deutsche Forschungsgemeinschaft (DFG), Grant Nos. SFB625/B4 and FOR597.
H.P. and A.M. acknowledge the use of computing facilities of Madara Computer
Center at Bulg. Acad. Sci.

%

\end{document}